# QoS-Aware Routing in Wireless Networks Using Aerial Vehicles

Vishal Sharma[1], Kathiravan Srinivasan[2]

[1] Computer Science and Engineering Department, Thapar University, India
[2] Department of Computer Science and Information Engineering, National Ilan University, Taiwan
vishal_sharma2012@hotmail.com, kathiravan@niu.edu.tw

## Abstract

The next generation wireless networks need efficient mechanisms for data dissemination that should support users with better Quality of Service (QoS). Nevertheless, the existing solutions are unable to handle this demand and require either network redeployment or replanning. Moreover, this upsurges the overall operational cost and complexity of the network. This problem can be addressed by deploying Unmanned Aerial Vehicles (UAVs), which can act as on-demand relays in next generation wireless networks. In this work, a novel strategy comprising a series of algorithms based on neural networks is devised, which resolves the issues related to data dissemination, QoS, capacity, and coverage. When compared with the existing methods, the proposed approach demonstrates better outcomes for various parameters, namely, throughput, message disseminations, service dissemination rate, UAV allocation time, route acquisition delay, link utilization and signal to noise ratio for end users. The experimental results exhibit the fact that the proposed approach utilizes 39.6%, 41.6%, 43.5%, 44.4%, and 46.9% lesser iterations than the EEDD, A-Star, OCD, GPCR, and GyTAR, respectively. Therefore, it is evident that the proposed approach surpasses the existing methods by means of superior performance and augmented efficiency.

## 1  Introduction

The next generation wireless networks aim at improving user experience in terms of coverage, capacity and Quality of Service (QoS). These networks facilitate a large number of users at higher data rates without causing any rendering effect on the performance of network components. The next generation wireless networks are all about the hybridization of components, which can be dynamically configured to provide better coverage and control over the entire network. One such collaborative dynamic network is formed by the incorporation of Unmanned Aerial Vehicles (UAVs) into existing networks that play a pivotal role in the selection and handling of User Equipment (UE), as shown in Figure 1.

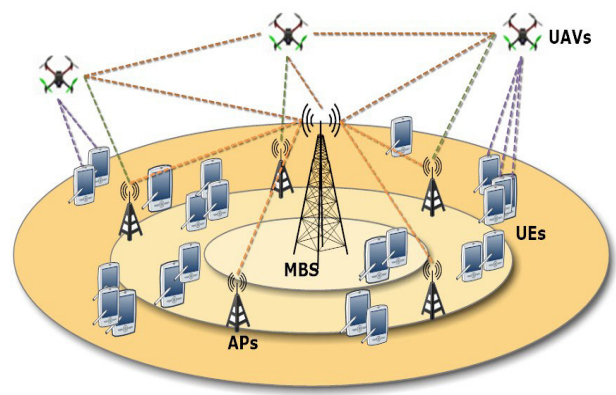

**Figure 1.** An illustration of the UAV assisted routing in wireless networks

The UAVs enhance the performance of existing wireless networks and resolve issues related to continuous transmissions. These aerial nodes can be used to overcome overheads, which may arise in a heterogeneous wireless network due to delay in handling users with high data rates [1]. Further, UAVs can be used to enhance connectivity in the public safety networks [2]. These vehicles reduce issues related to interference and provide high throughput coverage along with the improvement of spectral efficiency. These networks are capable of understanding the demand of users from a particular portion of a network governed by a macro cell for extra service support at similar data rates [3-4]. Further, UAVs provide versatility to these networks and are also able to resolve issues related to operation and maintenance of traditional wireless networks. The UAVs-assisted networks are also termed as "drone cell networks". Such networks are capable of providing better services than traditional networks because of their dynamic and easy to configure approach during network operations [5-7].

Traditional wireless networks consist of macro base stations (MBS), small cells, picocells, femtocells and UEs. According to architecture suggested under METIS, small cells, Radio Access Networks (RANs), Cloud-RANs, picocells, and femtocells form a crucial part of 5G deployment [8]. Also, massive machine communication (MMC) and device to device (D2D) communication are the crucial part of 5G scenarios. These components are the backbone of high-speed transmission in the next generation networks. However, deploying extra small cells, femtocells, and picocells to enhance the coverage increases complexity and cost of the overall network [5]. The proposed approach aims at providing efficient data dissemination without using the existing infrastructure of small cells, picocells and femtocells.

UAV oriented networks although provide a vast range of applications in the existing wireless networks, yet these come with a lot of challenges such as positioning of UAVs, allocation to demand area, QoS provisioning, and maintenance of route to facilitate a connection between UAVs, UEs, and MBS. The intensity of network is another major issue to be handled while operating UAVs since it directly affects the signal to noise plus interference ratio (SINR), which is a key metric in determining the performance of any wireless network. Thus, efficient approaches are required, which can consider the issues related to aerial coordinated network formations and can enhance the working of a traditional network without compromising with its performance, coverage and capacity [9-11].

In this paper, the problem of efficient data dissemination and QoS provisioning in next generation wireless network is considered. The problem deals with the initial mapping of UAVs to demand areas comprising some UEs, and then applying data dissemination approach to form a reliable network which is able to relay data efficiently in the case of indirect connectivity between UAVs, UEs, and MBS. There are some existing solutions, which also focuses on the same problem, but provides partial solutions only, i.e. either these solutions resolve the coverage and capacity issues or these only provide data forwarding in UAV-assisted networks such as distributed algorithm approach to optimally place UAVs for selection of an appropriate gateway [12]. This algorithm uses a UAV pattern division strategy to stabilize the UAV network.

Understanding the density of UAVs, velocity of UAVs, angle of arrival, and transmission range used by UAVs can also provide a strong support for handling the extra load in the next generation wireless networks [13]. But, this requires the understanding of load and identification of user demand areas. The control of trajectory and delay optimization can be a possible solution for efficient data dissemination, but control and alteration in course of UAVs require a lot network replanning and may result into non-serving of some crucial demand areas [14-15]. Resolution of the existing hardware can also help in improving QoS to some level unless efficient approaches are not used to take full advantages of these hardware changes [16].

The solution proposed in this paper utilizes neural network approach which forms priority sets over the key components of networks, and then uses series of algorithms to improve the working of existing networks. However, the proposed approach targets both the issues and presents an efficient hybrid solution which not only provides efficient data dissemination but also keeps intact the coverage and capacity of the network. The proposed approach is compared with some of the existing solutions to prove its effectiveness over standard network parameters, namely, throughput coverage, message disseminations, service dissemination rate, UAV allocation time, route acquisition delay, link utilization, and signal to noise ratio for end users. The proposed approach is evaluated for these parameters and compared with some of the existing solutions. The first one is energy-efficient data dissemination (EEDD) [17]. This approach provides a solution for data dissemination in UAVs-assisted wireless sensor networks. The results have been compared to test if this approach can be used for real-time traffic dissemination in next generation wireless networks. The second is A-Star [18], which provides data routing in the metropolitan vehicular networks. This approach is efficient for ad hoc networks. The comparative study analyzes it for UAV-assisted network environment. The third is opportunistic cross layer data dissemination (OCD) for flying ad hoc networks [19]. This approach uses a service layer abstraction to support data forwarding in flying networks. Fourth is the GPCR [20] algorithm, which provides routing support in the urban environments. This algorithm is also tested for its performance in the UAVs-assisted networks. The fifth is GyTAR [21], which provides an efficient greedy traffic aware routing for vehicular ad hoc networks. The proposed approach is tested against the existing solutions, and discussion is provided for the utility of existing as well as proposed approaches.

The remaining part of the paper is structured as follows: Section 2 presents the related work. Section 3 discusses the problem statement and our contribution. Section 4 presents the detailed system and network model. Section 5 gives a complete overview of the proposed approach along with detailed algorithms. Section 6 evaluates the proposed approach. Section 7 presents a comparison of the proposed approach with existing state-of-the-art approaches along with detailed discussion and open issues. Finally, Section 8 concludes the paper.

## 2 Related Works

The problem of data dissemination and QoS enhancement has been there for a long duration of time. However, dealing with UAVs, not much has been done towards the improvement of next generation wireless networks. Also, the existing approaches either work towards the betterment of capacity and coverage, data dissemination, or QoS [22]. None of the existing solutions aimed at all these aspects together.

In an existing work, Sharma and Kumar [23] have focused on the formation of an ambient network between the ground nodes and the UAVs to facilitate inter- and intra- UAV communication which aims at improving the quality of services to end users. The authors utilized neuro-fuzzy-genetic modeler approach to facilitate the information flow between network nodes. However, data dissemination is considered only as a part of intermittent connections and facilitation of ground users is not incorporated into their developed approach. Cortes et al. [24] have worked on the coverage control of mobile networks. Although their approach does not directly deal with the utility of UAVs, yet their approach uses a novel gradient descent algorithm which is capable of enhancing the coverage in autonomous networks. The utility of their approach to the autonomous aerial vehicles is still an open issue.

In a similar work, Hussein and Stipanovic [25] gave a coverage control mechanism for autonomous mobile networks with provisioning of collision avoidance. Although the approach developed by the authors is novel in functioning, but it does not focus on QoS and capacity of the next generation wireless networks. Facilitation of UAVs is also not included in their developed approach. Sharma et al. [5] have worked on the enhancement of capacity for heterogeneous networks. Their developed approach is capable of resolving a majority of parameters considered in this paper, but data dissemination along with provisioning of QoS is not included in their proposed coverage and capacity enhancement approach. In an extension of their work, Sharma et al. [26] suggested a proximity-sensitivity based routing for the data dissemination in UAV-assisted networks. The approach developed by authors help in finding an appropriate path in the UAV-assisted networks but does not provide any support for improving the quality of service and experience of end users.

Sharma et al. [17] also proposed an energy-efficient data dissemination approach for the UAV-assisted wireless sensor networks. Their approach is suitable for UAV oriented networks. This approach is presented especially for the sensor network formation considering UAV as the pivotal node. Despite limited domain, this approach can be implemented to UAV-assisted next generation wireless networks for capacity as well as QoS enhancement. This approach utilizes the properties of firefly optimization algorithm to select a path between the nodes. This approach is suitable for non-real-time data dissemination, similar to requirements of WSNs. However, in the next generation wireless networks, this dissemination has to be real time and should be swift enough that users enjoy high quality of experience throughout connectivity.

The solutions employed in existing vehicular technology can also provide some sort of remedy to the problem considered in this paper. However, these solutions cannot guarantee the performance until these are not tested over UAV scenarios. Some of the key solutions include, joint adaption approach by Rawat et al. [27], QoS-OLSR by Wahab *et al.* [28], and QoS guaranteed channel access approach by Chang et al. [29], and multi-constrained QoS aware routing by Eiza et al. [30]. A comparative study and evaluation of state-of-the-art approaches are presented in the later part of this manuscript.

The brief study of existing solutions suggests that novel approaches are required which can tackle the issues related to capacity, coverage, data dissemination and QoS provisioning together without affecting the other functionalities of a network. Thus, considering this as a problem, a novel approach is proposed in this paper which facilitates data flow in the next generation wireless networks using UAVs as key node along with provisioning of QoS.

## 3  Problem Statement and Our Contribution

Using UAVs in the next generation communication system is a tedious task. These aerial vehicles have to deal with a lot of issues related to their deployment and functionality. These vehicles can manage Capital expenditures (CAPEX) and operating expenses (OPEX) of existing heterogeneous networks and can extend their coverage and capacity [5]. However, the existing approaches available for device deployment as well as for UAVs are not capable enough to allow QoS to end users. Thus, efficient data dissemination and QoS provisioning along with enhancement of coverage and capacity is the main objective of this paper.

The proposed approach does not use or deploy extra small cells during data dissemination between UEs and UAVs. Thus, data dissemination is provided between UEs and MBS via UAVs. Here, UAVs act as the access point for UEs. The proposed approach uses a neural network approach to prioritize network requirements, and then uses a series of algorithms to efficiently disseminate data between network components. The proposed approach targets three paradigms, namely, data dissemination, capacity, and coverage to improve QoS in the next generation wireless networks.

## 4  System Model

The network comprises a set *M* of MBS covering an area *A* which is divided into *K* number of demand zones such that each zone comprises UEs making requests with an arrival rate of $\lambda$, such that $\sum_{i=1}^{K} x_i = |E|$. Here, *x* is the users in each demand zone and *E* is the set of UEs operating with a mean packet size $\frac{1}{\mu}$ which makes the network offered rate to be $\frac{\lambda}{\mu}$.

Further, the network utilizes a set $U$ of UAVs which facilitates UEs to support connections with MBS. The system model accounts for a link between MBS and UAVs, which facilitate the UEs with better QoS. The topology for MBS is done using a cell-based division, as shown in Figure 1. In the proposed approach, other components (small cell, picocells, and femtocells) of the HetNets are not considered in system modeling; since the network relies only on the connection between UAVs, MBS, and UEs. Also, the aim is to allow direct connection between MBS and UE using the intermediate UAVs as relays instead of small cells, picocells or femtocells. This connectivity allows a large number of users to be supported at the same instance with similar data rate and high signal quality. Each of these network components, UAVs, UEs, and MBS, share the same spectrum and aims at complete coverage with guaranteed throughput to most of the users. In the model, set $U$ of UAVs serves set $E$ of users with each UAV equipped with resources $R_c$ such that $R_c \leq |E|$. If a user in the network utilizes $R_e$ resources out of the available $R_c$ resources, then the total number of users handled in the network is given as:

$$\sum_{i=1}^{|U|} \left(\frac{R_c}{R_e}\right) \leq |E|. \quad (1)$$

For a consistent network, $R_e$ is same for all the UEs with same demand of the spectrum. Now, considering that a UAV can support multiple connections, the condition of load balancing $L_b$ for handling multiple users along with the connections with MBS and other UAVs is based on the number of uplinks available such that

$$L_b = T_u - (C_m + C_u), \quad (2)$$

where $T_u$ is the total users supported by a UAV, $C_m$ and $C_u$ represent the number of MBS and UAVs supported, respectively. Equation (2) can be utilized for load balancing in the case of UAV failures or in the case of a requirement for the extra facilitation of demand zones. Here, the connectivity between UAVs, MBS, and UEs is defined as the cost function of load which is to be minimized in order to allow the formation of a reliable network, such that considering the pending requests $R_p$, the cost function is given as:

$$C_f = \min(R_p), \quad (3)$$

where using [5],

$$R_p = \int_o^K \frac{\lambda}{\mu N \omega}, K \in A. \quad (4)$$

Here,

$$\omega = \beta \log_2(1 + SINR), \quad (5)$$

and

$$SINR = \frac{\frac{QW}{G^\alpha}}{\sum_{i=1, j \neq i}^{|U|} \frac{QW}{G^\alpha} + V_0}, \quad (6)$$

where $Q$ is the UAV transmission power, $W$ is a factor of antenna characteristics, $V_0$ is the thermal noise, and $N$ is the number of orthogonal bands into which the system bandwidth $\beta$ is split for the data rate $\omega$. Also, the Equation (4) is used to calculate the transmission delay while handling the pending requests marked by a derivative of the $R_p$ w.r.t. $A$. The complete network operates towards an increase in the capacity of UE during its connectivity with either a UAV or MBS. In a network, considering that the noise has a negligible effect on user experience, the per-UE capacity is given by:

$$U_c = \frac{\beta}{yN} \log(1 + SIR), \quad (7)$$

where $y$ is the number of users with $SIR$ below the threshold value. Now, using [31], the network intensity $\eta$ for a UE, considering negligible spectral density, is defined as:

$$\eta = \frac{c}{\pi G^2} \left(\frac{1}{SIR}\right)^{\frac{2}{\alpha}}, \quad (8)$$

where $G$ is the radio range between UAV and UE, $\alpha$ is the pathloss, $c$ is the network intensity constant which is dependent on the successful transmissions. Using Equations (8) in (7) [31], the per-UE capacity becomes:

$$U_c = \frac{\beta}{yN} \log\left(1 + \left(\frac{\lambda \pi G^2}{c}\right)^{\frac{-\alpha}{2}}\right). \quad (9)$$

Thus, from the network model, it can be observed that altitude of the UAVs will also play a key role in handling the UEs at higher data rates. Also, the level of intensity will also affect transmission capacity as well as the spectral efficiency of a network. However, these two aspects are not covered in this manuscript as it only focuses on the data dissemination, QoS provisioning by UAV to UE mapping, and the recovery mechanisms in case of UAV failures.

### 4.1 Network Model

The network model formed utilizes the underline system configurations to form a reliable data dissemination model which is capable of providing a better quality of service to end users. The network model for reliable data dissemination utilizes a reliability cost function $N_r$ to allow the formation of a stable and an efficient network such that

$$N_r \propto \frac{1}{S_d}$$
$$\propto O_c \quad (10)$$
$$\propto L_u.$$

Now, considering the normalizing constants for entire network, $\gamma_1, \gamma_2, \gamma_3 \in \gamma$ the reliability cost function becomes:

$$N_r = \max\left(\frac{\gamma_1}{S_d} + \gamma_2 O_c + \gamma_3 L_u\right), \quad (11)$$

where $S_d$ is the service cost with respect to deployed UAVs such that:

$$S_d = \frac{\sum_{i=1}^{|E|}(S)_i}{|U|}. \quad (12)$$

Here, $S$ is the service demand by each UE. $L_u$ is the link utilization cost which is given by:

$$L_u = \frac{N'}{N}, \quad (13)$$

where $N'$ is the number of bands utilized w.r.t. total bands $N$. The UAV utility cost $O_c$ is defined w.r.t. number of users $|E|$, if the maximum users handled are given by $|U|T_u$, where $T_u$ is the total users handled by a UAV, then

$$O_c = \frac{E_h}{|U|T_u}, \quad (14)$$

where $E_h$ denotes the actual number of users handled by UAVs. From $|E|$ users, some of the users are handled by UAVs, while some are handled by the MBS, such that $e_1+e_2=|U|q$. Here, $e_1$ is the number of users handled by UAVs and $e_2$ is the number of users handled by MBS. Now, the probability of users being handled by MBS or UAVs is given by:

$$\Pr(handled) = \frac{e_1 + e_2}{|E|},$$
$$= \frac{e_1 + (|U|T_u - e_1)}{|E|}, \quad (15)$$

and for Equation (15), $e_1+e_2 \leq |E|$. This UE to UAV mapping can be given as a likelihood $L_h$ of user being handled [32], with $n_0$ being allocated UAVs, such that

$$L_h = \max\left[\prod_{i=1}^{K \in A}\left\{\prod_{j=1}^{|E|}(\Pr(handled))^{n_0}(1-\Pr(handled))^{|U|-n_0}\right\}\right] \quad (16)$$

The maximization of Equation (16) allows complete mapping of UEs to UAVs such that all the users are handled efficiently at higher data rates.

The problem deals with efficient data dissemination between UAVs and UEs to support users at higher data rates. The network aims at provisioning of QoS to the maximum number of users apart from providing maximum coverage. The selection of optimized path between UEs and MBS via UAVs allow facilitation of a reliable network formation considering the cost functions defined in the network model. Thus, the problem deals with the formation of an efficient UAV-assisted data dissemination network considering the constraints of maximizing or minimizing the cost functions defined in Equations (3), (11) and (16).

## 5 Proposed Approach

The proposed approach aims at an efficient dissemination of data between UAVs and UEs along with enhancement of network coverage, capacity, and provisioning of QoS. The proposed approach operates in parts using multiple algorithms operating as individual threads such that entire model is handled in the form of a decisive approach using a neural schema. The neural schema is used to control the working of the proposed approach comprising multiple modules, where each module controls the network activity which results in the formation of a reliable network.

The neural network considered in this paper derives its working from the neural schema presented in Ref. [23] and aims at maximizing the network likelihood of mapping UAVs to UEs. This neural model operates on the priority of demand zones and the number of resources available with UAVs, which are fed back into the neural model to update after every iteration. The neural model forms a complete decision support system which is capable of controlling its learning, feedback, and error rate.

The learning rate defines the amount of alterations required in the allocation of UAVs to a particular demand zone. The feedback rate controls the flow of packets as information for the neural decision system, and the error rate identifies errors caused due to mismatch of UAVs and demand areas. The mismatch causes network model to reset for controlled activity. Alternatively, the error rate can be managed using a learning approach of this neural model which in turn enhances the network stability. The working of this neural model is further illustrated in Figure 2.

The neural model forms multiple priority sets P such that such that $P = P_1 \cup P_2 \cup P_3$, where $P_1, P_2, P_3$ defines the priority set of UEs, UAVs, and demand zones, respectively. The entire decision of the network depends on the dominance of these priority sets. In the considered approach, as data dissemination and QoS are the most important aspects, the priority order for these sets is considered as $P_3 > P_1 > P_2$, which suggests that the primary task is to handle demand zone initially, irrespective of the data rate, and if the demand zones are satisfied, attention is given to the QoS; and finally, the deployed UAVs are to be considered for their depleted resources or overconsumption. This schema allows decision making in the order of priority and allows efficient control over UAVs. Further, after abstracted control over the networks, the priority is also defined within these sets.

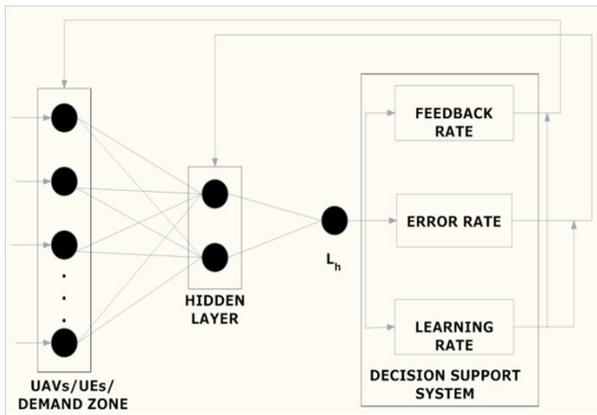

**Figure 2.** An illustration of the neural model

For the UEs, the priority order of the elements in the set $P_1$ is subjected to the number of requests generated by a UE and its distance from the MBS. Since, distance is a key player in the intensity of service offered, a UE with a large distance from MBS is unlikely to get any resource and must be handled with priority by the deployed UAVs. Thus, the priority order of $P_1$ elements will be subjected to the number of requests as a first parameter, and the distance as a second decisive parameter. In the case of similarity in these values, the priority is given to a UE whose demand zone is within the radio range of deployed UAV.

For UAVs, the elements of set $P_2$ are completely arranged on the basis of $R_c$, as defined in Equation (1). As a second parameter, $L_b$ is taken as a decisive parameter, as given in Equation (2). However, in the case of similar values, the coordinates of deployed UAVs are considered. The UAV with less distance from MBS is given the highest priority while taking any movement decision related to UAV reshuffling. However, the initial deployment is done on the basis of set $P_3$, which causes the entire network to follow a simple rule of first coming with a large request, gets served first with better resources.

### 5.1 UAV Allocation and Mapping to Demand Areas

The initial part of the proposed approach aims at the allocation of UAVs to demand areas so as to allow efficient localization of UAVs in the entire MBS zone.

The proposed model utilizes priority sets to map UAVs to demand areas. The demand areas with a priority higher than other are allocated UAVs with preference over the other zones. Also, the number of UAVs allocated to a single priority demand zone will be dependent upon the number of requests generated and the number of requests a single UAV can handle.

This procedure accounts for allocation of multiple UAVs to demand areas depending upon the requirement of enhanced coverage. The mapping of UAVs is another aspect of allocation. Once the UAVs are allocated to demand areas, mapping defines the number of simultaneous connections that a UAV will support to facilitate connectivity between MBS and UEs. The mapping is directly related to radio range of UAVs from demand zones.

The UAVs within the defined radio range will form a direct connection with UEs, whereas those allocated but not in the defined range utilizes the hop relaying to form a final connection with the UEs. This relaying can account for utilizing more UAVs within the transmission range of demanding UAV and demand area. The UAV allocation and mapping procedure are controlled by series of steps presented in Algorithm 1.

### 5.2 Mutual Peering and Control

The second part of the proposed approach deals with the mutual peering and control of UAVs for efficient coordination and voluntary exchange of traffic so as to allow maximum users in demand zones to be handled at higher data rates. The mutual peering allows multiple users to be supported by same UAVs despite being attended by other UAVs. The mutual peering takes into account the radio range of demand areas and UAVs. The initial allocation of UAVs to demand areas allow simple mapping between UEs and UAVs. However, because of area overlapping, these allocated UAVs will also accommodate other demand areas which are within their radio range.

Another aspect of this mutual peering is the control over network. Usually, the MBS is a controller of the entire network and so as the UAVs. The network is supported by the decisions of MBS made on the basis of demand requests. However, for further enhancement of connections, the UAVs with radio range less than equal to the permitted range will allow command and control over other UAVs. This UAV will account for a check on failed UAVs which will be discussed in the next part. The entire procedure for mutual peering and control is presented in Algorithm 2.

### 5.3 Failure Control and Load Balancing

Network operating with dynamic nodes is liable to have some sort of failures which may be accounted in the form of UAV failure, MBS failure or some other connection loss. These failures may hinder the network operations and can cause network shutdown. However,

**Algorithm 1.** UAV to UE allocation and mapping

1: Input: U, K, A, E, G
2: Initialize the Network
3: MBS= Check demand zone announcements
4: while (All areas are not covered) do
5:    compute K
6:    $K_u$=calculate the number of UAVs required by K zones
7:    Find priority $P_3$ order of the demand zones
8:    if (priority defined ==true ) then
9:      Arrange K in Descending order
10:      i=1
11:      while (i ≤ K) do
12:        j=1
13:        while (j ≤ max($K_{u,i}$)) do
14:          Allocate UAVs to K
15:          j=j+1
16:        end while
17:      D=Compute radio range of allocated UAVs
18:      if (D ≤ max(G)) then
19:        Map UAVs and start transmission
20:      else
21:        Reset and repeat steps 3 onwards
22:      end if
23:      i=i+1
24:     end while
25:   else
26:     calculate error rate
27:     set feedback and learning rate and Update neural schema
28:     Reset and repeat steps 3 onwards
29:   end if
30:   continue
31: end while
32: Mapping Successful

**Algorithm 2.** Mutual peering and control

1: Input: U, K, A, E, G
2: Initialize the Network with the allocated UAVs
3: while (Transmission!=complete ) do
4:   Check for allocated UAVs ← U
5:   i=1
6:   while (i ≤ |U|) do
7:     Mark current radio range D[i]
8:     if (D[i] ≤ G) then
9:       accept connections from demand areas
10:     else
11:       continue with the allocation
12:     end if
13:     $C_u$=select the UAV with least distance from MBS
14:     Mark $C_u$ as the controller
15:     i=i+1
16:   end while
17: end while

**Algorithm 3.** UAV failure and load balancing

1: Input: U, K, A, E, G
2: Initialize the Network with the allocated UAVs
3: while (Transmission!=complete ) do
4:   Check for allocated UAVs ← U
5:   Mark any unresponsive UAVs
6:   $U_d$=Check for the unhandled demand area
7:   if (UAV ↔ $U_d$) then
8:     Check for the requirement of extra UAVs
9:     if (Requirement == true) then
10:       launch new UAVs
11:     else
12:       continue with the pre-allocation
13:     end if
14:   else
15:     $U_u$=account for underload UAVs
16:     i=1
17:     while (i ≤ |$U_u$|) do
18:       Calculate $L_b$[i]
19:       i=i+1
20:     end while
21:     allocate UAVs with $L_b$ ≥ 1 to the demand area
22:     re-route traffic to perform load balancing
23:   end if
24: end while

reliable networks should be capable of handling these failures and should be able to re-configure themselves to overcome these failure issues. Further, this control over failure would require provisioning of load balancing so as to allow continuous flow of traffic.

The part considered in this paper deals only with UAV failures and accounts for an alternative load balancing approach in order to overcome the failures. The failure control is performed with the help of controller node selected using Algorithm 2, and then the MBS and allocated UAVs takes a collaborative decision on the load balancing of the entire network. The procedure for the load balancing is handled using Equation (2) and the steps are presented in Algorithm 3.

### 5.4 Data Dissemination and QoS Scheduling

The network approach proposed in this paper aims at provisioning of QoS to end users despite the network demands and connections. This provisioning of QoS is facilitated by UAVs which help in balancing the load as well as undergo data dissemination procedures to select an optimal route in the case of multi hop relaying. This data dissemination is required in the case of indirect linking between UAVs, UEs, and MBS. For direct linking, UAVs utilizes the mapping procedure defined in Algorithm 1.

This mapping is capable of enhancing the coverage of network but up to some extent as the direct linking is totally based on the radio range and UAVs are to be connected with MBS at all instance. However, an alternative multi-hop relaying between UAVs in the next generation wireless networks will add an extra paradigm which will not only facilitate more UEs but will be able to handle extra network load without any hindrance. However, this indirect linking although increases the coverage but affects data rate. Thus, an efficient procedure is required which can handle these dynamic route selection and also incorporate the dynamic load balancing to increase network capacity without affecting the data rate. Also, all these procedures are to be done with minimum route acquisition delay as this would directly affect the quality of service. The entire procedure for data dissemination is controlled by the parameters, namely, $G$, $L_b$, $|U|$, $N_r$, the location of UAVs, and the number of UEs currently handled by each UAV. The proposed data dissemination is further categorized into three main aspects, namely, route selection, route rehabilitation, and route maintenance.

**Route selection.** The initial phase of UAV allocation to demand area is marked by a simple mapping procedure. This mapping is then accounted for network reliability, which forms a threshold value $N_r^{TH}$. This threshold value is used in the selection of routes in the case of indirect linking between UAVs and MBS considering the radio range of UAVs and UEs. The route selection also takes into account the network intensity $\eta$ which provides a range of users with guaranteed SIR. The procedure for route selection is presented in Algorithm 4. This algorithm utilizes hello messages to identify UAVs operating on indirect linking. On the basis of these hello messages, a routing table is formed which marks the entries comprising $G$, $\eta$, $L_b$, $N_r$, and the location of UAVs. With each iteration or addition of extra UAVs in the route, all these metrics are checked again and compared with the threshold value to allow selection of an optimal route with high data rates and less route acquisition delay. Since the entire algorithm is operated over each UAV, a continuous check is performed on the connections which prevent the network from idle state and also allows efficient load balancing.

**Route rehabilitation.** Route rehabilitation is the part of the proposed approach which aims at correction of routes in case of failures. Route rehabilitation also aims at load balancing in the network. In the case of multiple component failures, load balancing regulates the traffic flow without affecting the initial data rate. The route rehabilitation derives its complete functionality from the steps defined in Algorithm 3. The route rehabilitation also takes into account $N_r$ as it would allow selection of a correct and an optimal route

**Algorithm 4.** Route selection
1: Input: U, K, A, E, G, $N_r^{TH}$, $L_b$
2: Initialize the Network with the allocated UAVs and check for HELLO messages
3: while (Transmission!=complete ) do
4:    Select the UAVs with indirect connections
5:    P2 =Mark the UAVs in order of their priority
6:    Check for the location of each UAV
7:    Select the UAVs with defined radio range $\leq$ G
8:    Arrange UAVs in order of $L_b$
9:    X=count (Arrange UAVs in order of $L_b$ )
10:    j=1
11:    while (j $\leq$ X) do
12:      mark each UAV and calculate N j
13:      if ($N_r^j \leq N_r^{TH}$) then
14:        Add X in path[]
15:      else
16:        skip the UAV
17:      end if
18:      j=j+1
19:    end while
20:    select shortest path on the basis of distance and G
21:    transmit
22: end while

for every iteration. The alteration in algorithm is performed at step 21 which also checks for $N_r \geq N_r^{TH}$ during route rehabilitation.

**QoS maintenance.** The quality of service is a major demand of the next generation wireless networks. Providing QoS and SIR to UEs in the heterogeneous networks are major requirements of these networks along with the enhancement of network capacity and coverage. The QoS of network is subjected to constraints defined in the Equations (3), (11) and (16). These equations check load, network reliability and network likelihood for handling the users at higher data rates. QoS provisioning is a complex task in a network with dynamic nodes as these components may change their operating behavior during network operations. However, controlling these components and selecting the best suitable dynamic node can resolve this issue and can guarantee higher QoS to end users despite their location and geographical distance. The proposed approach utilizes three constraints to check QoS and to select an alternative route in the case of failure in meeting any of these selection criteria.

The proposed approach continuously checks for these constraints and selects an alternative optimal path with second best values for these parameters as a route to transmit data in the case of indirect connection between UAVs and UEs, or between UAVs and MBS. These steps for this QoS maintenance are provided in Algorithm 5. This algorithm takes the current state of network as an input and then decides on the constraints to check for conditions which are not satisfied during route selection and data dissemination. These conditions are then used to alter the current state of network, and finally provide an updated path which can be used to regulate data transmission at higher rates. The threshold values for these constraints are taken from their values during the initial mapping of UAVs with UEs depending upon the decision of MBS. These thresholds can be altered depending upon the requirement of network connections. In the case of extensive converge requirement, these values can be relaxed and the network can be directly used without any updates for $N_r$, $C_f$, and $L_h$.

---

**Algorithm 5.** QoS Maintenance

1: Input: Current State ← U, K, A, E, G, $N_r^{TH}$, $C_f^{TH}$, $L_h^{TH}$
2: Fetch results from Equations (3), (11), (16)
3: while (Transmission! =complete) do
4:     select the current route []
5:     check for the $N_r$, $C_f$, $L_h$
6:     set priority for constraints
7:     if ($N_r \geq N_r^{TH}$ && $C_f \geq C_f^{TH}$ && $L_h \geq L_h^{TH}$) then
8:         continue with the selected route
9:     else
10:        reset the network and re-initiate the route selection
11:    end if
12: end while

---

### 5.5 Decision Model and Flowchart

The proposed approach deals with the allocation of UAVs, mutual peering, control failures, data dissemination, and QoS scheduling by incorporating the defined algorithms. The proposed network model deals with these issues by incorporating a series of algorithms which provide a stable and a reliable network formation which not only improves the coverage but also improves the user experience by selection of an optimal path between multiple UAVs and UEs. This path selection is subjected to the condition of direct or indirect connections. For the direct connections, UAVs mapped to UEs forms the threshold conditions which are then checked during inter-UAV relaying. This helps in the selection of a reliable and an optimal path which can provide high data rate services to the entire network zones.

The decisive model incorporates all other algorithms defined in above procedure for taking decisions related to final route selection and QoS based data dissemination. A detailed flow chart of the complete approach is defined in Figure 3. This flowchart helps to understand the evaluation and implementation strategy of the proposed approach. The flow chart suggests that first of all MBS is initialized, which will regulate the network flow, and then the network is marked with segments. After segmentation, the position of UAVs is initialized with an identification of UEs and their demand zones.

Next, the neural model forms the multiple priority sets as defined in Section 5. The entire decision of network depends on the dominance of these priority sets. In the considered approach, as data dissemination and QoS are the most important aspects, the priority order for these sets is considered as $P_3 > P_1 > P_2$, which suggests that the primary task is to handle demand zone irrespective of its data rate, if demand zones are satisfied, attention is given to QoS, and finally, the deployed UAVs are to be considered for their depleted resources or overconsumption. Next, the UAVs are assigned and a route is selected for transmission with the maintenance of QoS as well as the shortest path. At any point, leading to dissatisfaction with the underlying conditions, the network is reset to neutral phase, and priority sets are recomputed before proceeding further.

It is to be noted that the flight dynamics for UAVs are not considered, while development of the proposed approach. The UAVs are to be deployed over demand area with maximum requests. This deployment allows an efficient positioning of UAVs. Only the initial path between a source and the destination is selected on the basis of radio range, but after the establishment of the first link, this path is updated to allow selection of the route, which guarantees optimal QoS to UEs, as shown in Algorithm 3, 4, and 5. Thus, the selected path is capable of allowing optimal throughput to UEs. However, there is a constraint, while considering the coverage scenario. Since, the coverage requires a shift of services from UAVs to UAVs; this raises an important issue of handovers in the UAVs-assisted networks. This leads to inclusion of new or existing handover features to resolve communication overheads related to the shift of services. This will certainly affect the service timings, which may cause extra delays in the network. Also, extra communication equipments or protocols have to be added to resolve this issue. Nevertheless, this issue is not directly targeted in the proposed approach, and will be studied independently in the future work.

Apart from handovers, maintenance of backhaul link is another critical aspect of UAVs-assisted next generation wireless networks. The UAVs in these networks are capable of providing a continuous backhaul support using the existing wireless technology. All the deployed UAVs are continuously connected to their parent MBS and share control information with them. This connection is similar to the one provided between the access points and base station in traditional wireless networks.

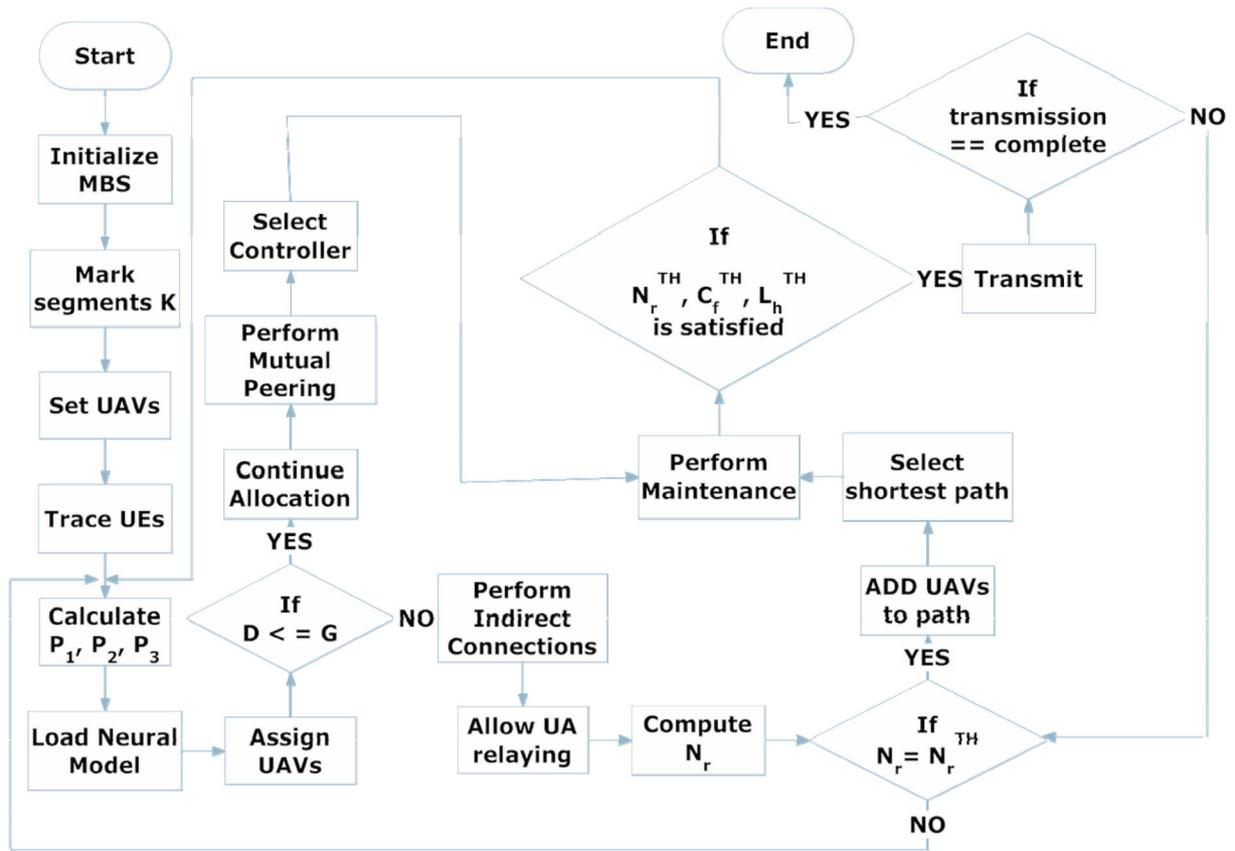

**Figure 3.** Flow chart for complete data dissemination using UAV

## 6 Performance Evaluations

The proposed approach was analyzed using network simulations performed in the scenario generated using MATLAB. The network was analyzed over an area of 10000 × 10000sq.m. with 10 MBS per sq. km. Each of the MBS zones is subdivided into 12 segments which possess some UEs dynamically allocated to generate 2 requests per second. These segments are the demand zones into which a single cell is divided to efficiently localize the UE. The system model in Section 4 accounts for a link between the MBS and the UAVs which facilitates the UEs with better QoS. The topology for the MBS is done using a cell- based division, as shown in Figure 1, such that each of the ground units is capable of communicating with every connected device using regulations of METIS [8]. For UAVs, the segments/demand zones serve as the positioning system for UAVs. Since, line of sight plays a key role in the UAV-assisted networks; a mesh topology is formed between the UAVs maneuvering a single cell. The remaining configurations for network simulations are presented in Table 1.

**Table 1.** Parameter configurations

| Parameter | Value | Description |
| --- | --- | --- |
| A | 10000x10000 sq. m. | Simulation Area |
| \|M\| | 10 per sq. km | Number of MBS |
| K | 12 (per MBS) | Segments per MBS |
| \|U\| | 1-10 (per MBS) | Number of UAVs |
| G | 500 m | Radio Range |
| μ | 256 kbps | Offered Traffic |
| α | 4 | Path loss Exponent |
| β | 10 MHz | System Bandwidth |
| \|E\| | 100-1000 | Active Users |
| $max(C_m)$ | 5 | Connections with MBS |
| $max(C_u)$ | 5 | Connections between UAVs |
| N | 2-5 | Number of bands |
| S | 2 per second | Service requests |
| W | -11 dB | Transmission Constant |
| Q | 35 dBm | Transmission Power |

The proposed approach was evaluated for various parameters and compared with the energy efficient data dissemination (EEDD) [17], A-Star [18], Opportunistic cross layer data dissemination (OCD) [19], GPCR [20] and GyTAR [21]. The detailed evaluations, results and discussions are presented below:

### 6.1 Throughput Coverage

Initially, the network was analyzed for throughput coverage provided by the proposed approach. Throughput coverage is the percentage of users covered with higher data rates during network transmissions. The proposed approach utilizes multiple algorithms to facilitate UEs in the zone of MBS with higher data rates. The proposed approach not only allocates UAVs to the demand areas but also provide high data rate to each UE. This is analyzed in terms of percentage throughput coverage. A network with more throughput coverage is capable of providing better capacity as well as coverage in terms of the number of users with high data rates. The analysis presented in the paper suggest that the proposed approach is capable of providing better throughput coverage than the existing approaches even with an increase in the number of UEs. Further, the mutual peering and controller selection also facilitates network throughput and provides vast coverage without any hindrance. The results presented in Figure 4 show that the proposed approach provides 5.7%, 7.3%, 9.03%, 10.1%, 13.3% better throughput coverage than the EEDD, A-Star, OCD, GPCR, and GyTAR, respectively.

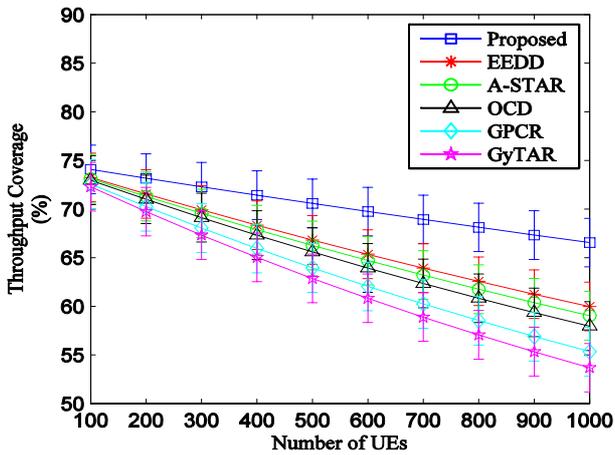

**Figure 4.** Throughput coverage vs. UEs

### 6.2 UAV Allocation Iterations

The procedure for efficient data dissemination is initiated by the allocation of UAVs to demand area. After allocation, the route selection procedures are used to select an optimal route. However, allocation of the UAVs to demand areas requires less iteration to optimize network formations. The number of iterations required to map UAVs to the respective area is one of the crucial tasks in the next generation wireless networks.

Although the existing approaches do not directly provide this facility of mapping UAVs to demand area, yet the allocation of UAVs to required area on the basis of location is performed in order to allow their comparative analysis with the proposed approach. The proposed approach explicitly provides a facility of mapping UAVs to demand areas with a lesser number of iterations as suggested in Algorithm 1. The algorithm iterates only if there is an error in the network mapping. Further, the iterations required by this algorithm is added in the form of processing delay in calculation of the end to end delay. The analysis presented in Figure 5 illustrate that the proposed approach requires a large number of iterations to map UAVs to demand area, but this increase is sufficiently lower than the existing solutions. The result concluded that the proposed approach utilizes 39.6%, 41.6%, 43.5%, 44.4%, and 46.9% lesser iterations than the EEDD, A-Star, OCD, GPCR, and GyTAR, respectively.

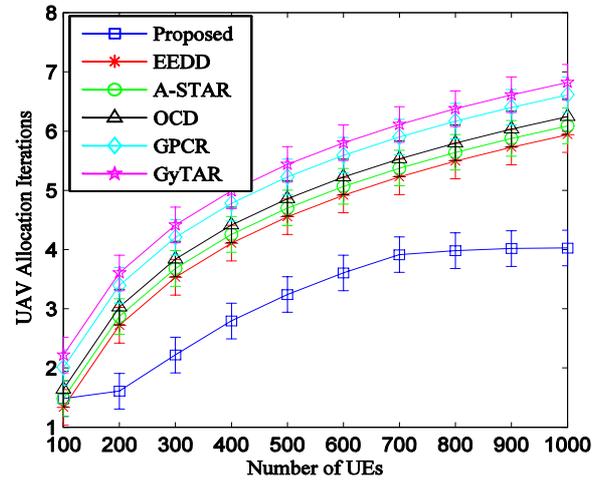

**Figure 5.** UAV allocation iterations vs. UEs

### 6.3 Percentage Users with Guaranteed SIR

The signal to interference ratio is another measure of the network performance. A network which is capable of providing better SIR to maximum users is capable of forming a reliable connection which can facilitate the maximum number of users at higher data rates. The percentage of users with guaranteed SIR is the measure of users with SIR above a certain threshold value. For the analyses, this threshold is considered equal to the SIR at an initial allocation of UAVs to demand zones when all the UEs are handled by the MBS. The allocation of UAVs increases the coverage and capacity of the network providing better SIR to maximum users. However, with an increase in UEs, this percentage also decreases as the network resources remain same despite the increase in UEs. However, the proposed approach allows reshuffling and resetting of network state in the case of increase in UEs and maintains a check on the network constraints which allows maximum users to be served with better SIR. The analyses presented in Figure 6 show that the proposed approach covers 2%, 2.3%, 7.1%, 14.2%, and 16.6% more users with guaranteed SIR than the EEDD, A-Star, OCD, GPCR, and GyTAR, respectively.

### 6.4 Per UE Capacity

The network which aims at the formation of a reliable connection between users and dynamic nodes must provide a better capacity to its user. This capacity is measured in terms of per UE capacity which allows analysis of the transmission capacity offered by the network to each user being mapped by a UAV. The proposed approach maps UEs to allow coordination between the users and the MBS in order to enhance the number of users by handling their service demands. The proposed approach is capable of providing a better probability for users of having high per UE capacity in comparison with the existing approaches. The analyses presented in the Figure 7 show that the proposed approach is capable of enhancing the per UE capacity of network considering the deployment using Algorithms 1 and 2, such that the proposed approach allows an improvement of 2.8%, 7.1%, 17.1%, 15.7%, and 14.2% in comparison with the EEDD, A-Star, OCD, GPCR, and GyTAR, respectively.

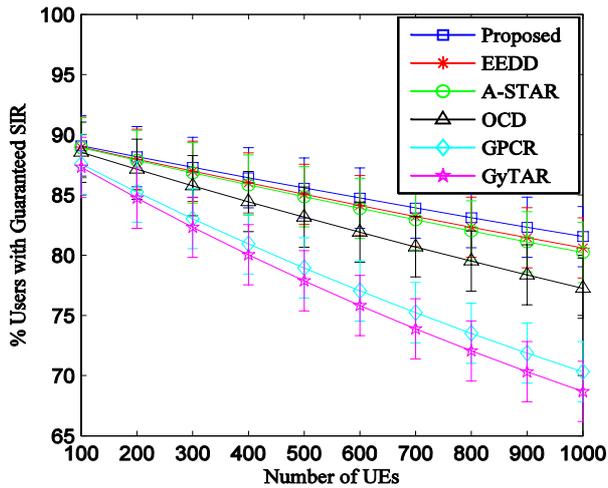

**Figure 6.** Guaranteed SIR to users vs. UEs

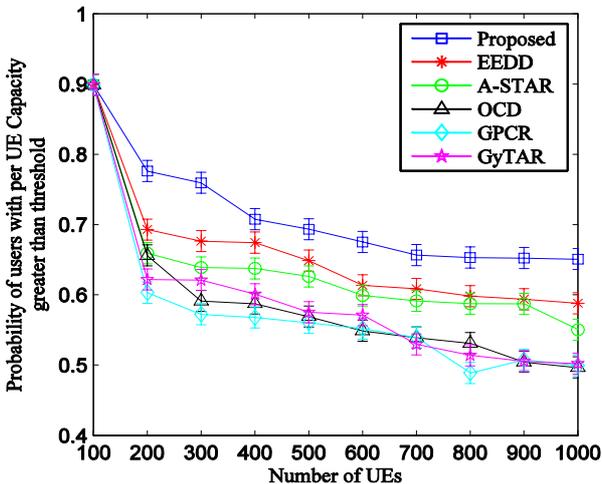

**Figure 7.** Probability of per UE capacity vs. UEs

### 6.5 Message Disseminated

The major role of a data dissemination approach is to allow better message servicing between the users of a network. The number of messages disseminated is a measure of the delivery ratio of the network which accounts for successful message transmissions to the total message generated and transmitted over the network. A network with better message dissemination is better in terms of data forwarding and delivering between its users. The percentage of message disseminated by the proposed approach increases with an increase in the number of UEs, as more messages float in the network with more UEs. The initial configuration of the network suggests that each of the active users make service requests of at least 2 units per second which increase as the number of UE increases. The proposed approach is capable of sustaining this increase and provides a better percentage of messages disseminated even with an increase in the number of active users. Figure 8 presents the comparative plot for percentage message disseminated over the network and shows that the proposed approach provides 3.9%, 6.5%, 8.7%, 10.0%, and 11.1% better message dissemination than the EEDD, A-Star, OCD, GPCR, and GyTAR, respectively.

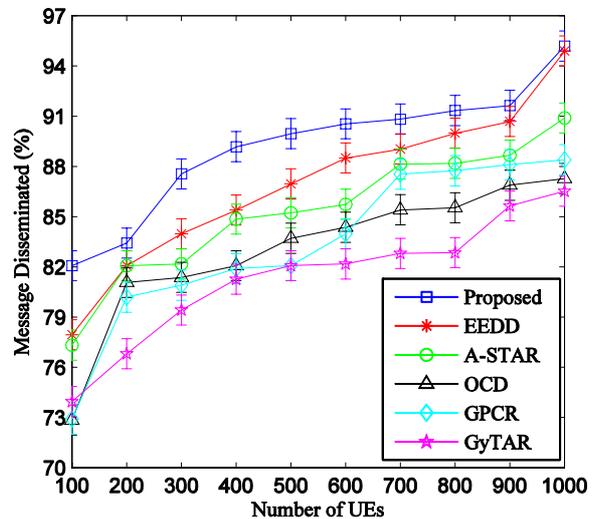

**Figure 8.** Message disseminated vs. UEs

### 6.6 End to End Delay

A network aiming at better QoS along with the efficient data dissemination must be able to support the entire network services with lower end to end delays. The end to end delays is the measure of transmission delay, propagational delay, processing delay, and queuing delay imposed during the transmission procedure. Here, processing delay accounts for running time of each algorithm used for mapping UAVs to UEs. With an increase in the number of UEs, the delays are bound to increase, but this increase should not hinder the network performance and should be less enough that the other network services are not affected by it. The proposed network approach provides efficient route selection, rehabilitation and maintenance facility which prevents the network from delays during the entire procedure, thus, causing the network to show an improvement of 25%, 40%, 50%, 62.5% and 69.3% in comparison with the EEDD, A-Star, OCD, GPCR, and GyTAR, respectively, as shown in Figure 9. The lesser delay allows the formation of an efficient network with better QoS to end users throughout the connectivity and network operations.

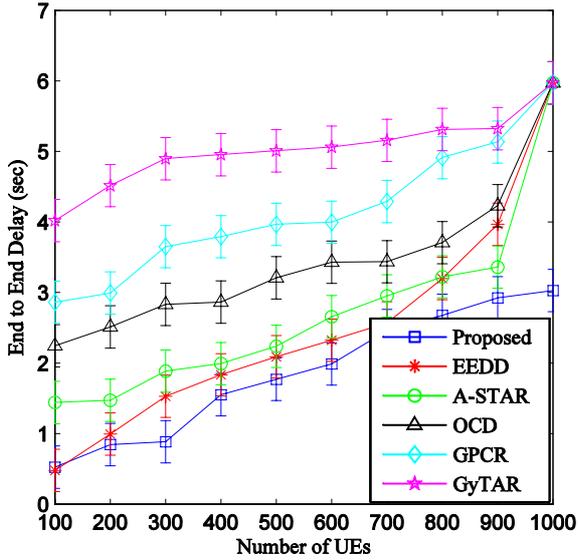

Figure 9. End to end delays vs. UEs

### 6.7 Link Utilization

The enhancement in service quality is embarked by the improve- ment in link utilization of the entire network. Link utilization is dependent upon the number of UAVs deployed to facilitate UEs in the demand areas. The system model is formulated for UAVs rather than MBS, which operates on an orthogonal band. Further, UEs operate with different QoS requirements, while UAVs are operated over the same frequency band i.e. UAVs do interfere and the model defined in (5) and (6) takes care of this interference. Here, link utilization refers to the number of bands utilized by the entire network of UAVs, i.e. with more UAVs, more bands are available, and thus, link utilization increases despite an increase in the number of UEs. The percentage utilization of the link is based on the active channels supported by dynamic network components. In the proposed approach, these dynamic network components are UAVs which are capable of supporting multiple UEs with better link quality and band utilization, thus, increasing the overall percentage of link utilization by 14.2%, 18.3%, 23.4%, 28.5%, and 33.6% in comparison with the EEDD, A-Star, OCD, GPCR, and GyTAR, respectively, as shown in Figure 10.

### 6.8 Service Dissemination Rate

The number of services offered per second by the UAVs to requesting UEs is the measure of service dissemination rate. The service dissemination rate allows checking service support for the entire network. A network with better QoS allows higher service dissemination rate which causes a number of user requests to be handled at the same instance, thus, increasing the coverage, capacity, and QoS simultaneously. This metric is used to measure the network performance during entire session of connectivity between the deployed UAVs and the demanding UEs. With an increase in the number of UAVs, more bands are available; also the connectivity between UEs, MBS and UAVs increases with an increase in the number of these aerial vehicles causing the large increase in the number of services being handled at the same instance. The analysis presented in Figure 11 show that the proposed approach is capable of providing 6.75%, 13.25%, 26.25%, 25%, and 16.25% better service dissemination rate than the EEDD, A-Star, OCD, GPCR, and GyTAR, respectively.

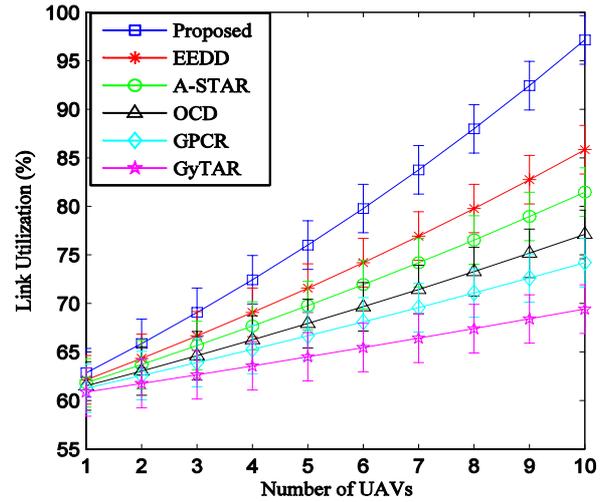

Figure 10. Link Utilization vs. UAVs

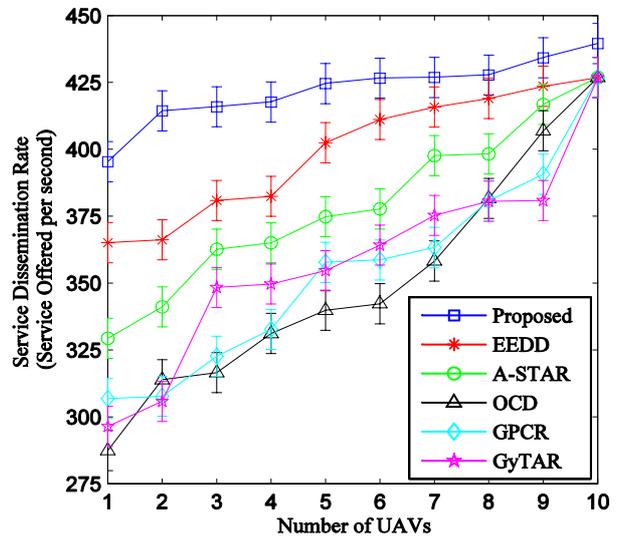

Figure 11. Service dissemination rate vs. UAVs

### 6.9 Route Acquisition Delay

A network operating with multiple nodes is liable to undergo multi-hop relaying. The proposed approach facilitates both direct as well as indirect connectivity between the UAVs, UEs, and MBS. The proposed approach allows the selection of an optimal route by the formation of priority sets using a neural schema provided the constraints on $L_b$, $C_f$ and $L_h$ are satisfied by the selected route. The proposed approach further enhances the selected route by providing load balancing in the case of network failures. However, this entire procedure is dependent on the selection of a route after certain iterations which may cause sufficient delay to hinder the network performance. A network undergoing consistent route changes must have low route acquisition delay so as to allow efficient data dissemination without affecting the QoS. The proposed approach selects a route on the basis of constraints which are checked in a single pass, thus, allowing routing to be done with lesser delay. The comparative plot in Figure 12 show that the proposed approach cause 3.8%, 9.6%, 16.6%, 27.8%, and 31.8% lesser delay in final route acquisition in comparison with the EEDD, A-Star, OCD, GPCR, and GyTAR, respectively.

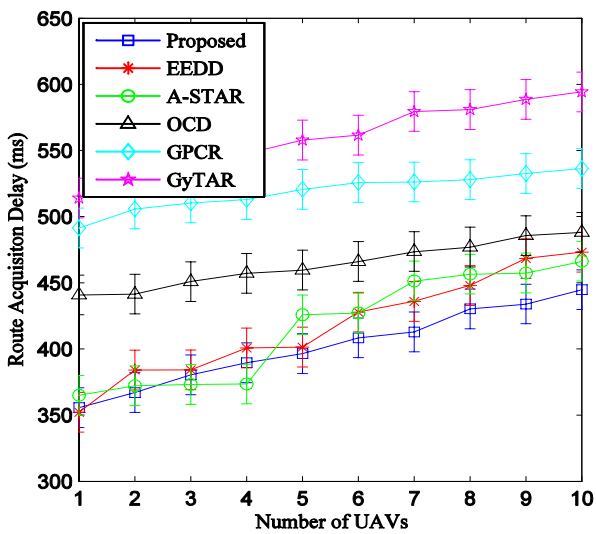

**Figure 12.** Route acquisition delay vs. UAVs

The summarized results demonstrating the percentage improvement in the proposed approach in comparison to the existing solutions is shown in Table 2. The analysis presented in this paper show that the proposed approach is capable of providing an efficient strategy for data dissemination which not only marks the UAVs to particular demand area but also supports QoS provisioning for the end users. The proposed approach utilizes a series of algorithms which manages the flow of traffic and selection of appropriate UAVs in the case of indirect connectivity between the UAVs, UEs, and MBS. The proposed approach improves coordination of the next generation wireless networks by using UAVs as its key node and also facilitates its users with higher data rates.

## 7 State-of-the-art Comparison, Discussions and Open Issues

The proposed approach is evaluated against various existing approaches, namely, EEDD, A-Star, OCD, GPCR, GyTAR, as shown in the previous section. Apart from these comparative evaluations, the proposed approach is also evaluated against the existing state-of-the-art solutions, as shown in Table 3. Although these approaches are designed for UAVs networks, but these only provide a solution to some of the issues which are considered in this paper. Due to this, only a tabular comparison is presented rather than the graphical layouts. The comparison has been drawn on the basis of support and solution for particular parameters over which the proposed approach is evaluated.

The comparison shows that most of these approaches have focussed only on the coverage and have not presented a strong solution to data/service dissemination in UAVs-assisted next generation wireless networks. Further, the comparison also suggests that the proposed approach deals with more number of parameters, whereas the existing approaches have focused only on the limited number of parameters, thus, providing a partial solution to the considered problem. Also, the other literature so far has focussed on the ad hoc formations between the ground vehicles and has treated UAVs similar to these vehicles. This is not an efficient and correct approach as the UAVs are having altogether different maneuvering capabilities and are relatively more dynamic. Also, these aerial vehicles require more network management than the ground networks, thus, making most of the existing work of ground network inapplicable in these networks.

The comparison also suggests that approaches in the UAVs-assisted network should handle all types of tradeoffs to allow better utilization of network resources. Also, the primary focus should be on the improvement of user experience by facilitating network with low overheads and low delays. Apart from these, the proposed approach should be scalable to allow the addition of more network devices and UAVs without changing the network configurations. The comparison shows that the proposed approach can withstand a large number of devices and UAVs without any requirement of extra UAVs and network-replanning.

**Table 2.** Average Improvement in the proposed model

| Parameter | EEDD | A-Star | OCD | GPCR | GyTAR |
|---|---|---|---|---|---|
| Throughput Coverage | 5.70% | 7.30% | 9.00% | 10.10% | 13.30% |
| UAV Allocation Iterations | 39.60% | 41.60% | 43.50% | 44.40% | 46.90% |
| Guaranteed SIR | 2% | 2.30% | 7.10% | 14.20% | 16.60% |
| per UE Capacity | 2.80% | 7.10% | 17.10% | 15.70% | 14.20% |
| Message Disseminated | 3.90% | 6.50% | 8.70% | 10.00% | 11.10% |
| End to End Delay | 25% | 40% | 50% | 62.50% | 69.30% |
| Link Utilization | 14.20% | 18.30% | 23.40% | 28.50% | 33.60% |
| Service Dissemination Rate | 6.75% | 13.25% | 26.25% | 25% | 16.25% |
| Route Acquisition Delay | 3.80% | 9.60% | 16.60% | 27.80% | 31.80% |

Table 3. Comparison of the proposed approach with state-of-the-art models: High (H), Medium (M), Low (L)

| Approach | Author (Year) | Throughput coverage | Message dissemi-nation | Per UE capacity | E2E delay | Service dissemi-nation | Link utilization | Routing overheads | Scalability |
|---|---|---|---|---|---|---|---|---|---|
| UAV Assisted Delay Optimization | Sharma et al. (2016) [1] | - | - | - | L | - | - | L | H |
| UAVs Assisted Networks for Public Communication | Merwaday and Guvec (2015) [2] | H | H | - | L | - | - | - | M |
| Efficient 3-D placement of an aerial base station | Bor-Yaliniz et al. (2016) [15] | M | - | M | - | H | - | - | H |
| Triple-band printed antenna array | Khawaja et al. (2016) [16] | - | - | - | - | - | H | - | H |
| Dynamic Trajectory Control Algorithm | Fadlullah et al. (2016) [14] | - | - | - | L | H | H | L | H |
| Enhancing Connectivity for UAV Communication | Rawat et al. (2015) [13] | - | M | - | L | - | M | - | M |
| Gateway Selection Algorithm | Luo et al. (2015) [12] | - | - | - | L | - | - | L | H |
| Proposed Approach | Sharma and Srinivasan | H | H | H | L | H | H | L | H |

Apart from the solution provided in this work, there are several other issues which are to be taken care of while dealing with UAVs-assisted wireless networks. Some of these include, efficient strategy for handover mechanism in single as well as in group mode, efficient approaches resolving issues related to battery optimization of these aerial vehicles, user-facilitation and choice of network selection, capacity and spectral efficiency can further be improved. Apart from these, there are several issues related to security in these assisted networks such as UAV hijacking, UAV trapping, UAV-query-manipulation, etc. All these issues must be resolved to form fully functional, reliable and secure UAVs-assisted next generation wireless networks.

## 8  Conclusion

In this paper, data dissemination along with QoS provisioning in UAV-assisted next generation wireless network was considered. The proposed approach utilized the feature of priority based neural network that allows the selection of an optimal path between the UAVs, UEs, and MBS in the case of indirect mapping without affecting the coverage and capacity of the considered network. The issues of a traditional network comprising small cells with MBS are overcome in this work. The proposed approach proves to be efficient in the formation of an efficient network with higher data rates to the end users.

The paper first provides the system and network models for the UAVs network. Then, it discusses the proposed work, which utilizes the neural network to resolve issues related to mapping of UAVs and UEs by forming priority sets. Then, it uses a series of algorithms to provide support for data dissemination along with mutual peering. Subsequently, the proposed approach is evaluated against the existing approaches and comparison is drawn with state-of-the-art models to prove its efficiency and superior performance.

A section on discussion and open issues is also presented, which provides insight to some of the key problems in these types of networks. The results presented in this paper show that the proposed approach offers better throughput, message disseminations, service dissemination rate, UAV allocation time, link utilization, signal to noise ratio for end users and lower route acquisition delay in comparison with the energy-efficient data dissemination (EEDD), A-Star, Opportunistic cross layer data dissemination (OCD), GPCR and GyTAR.

## Biographies

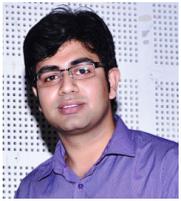

**Vishal Sharma** received the Ph.D. and B.Tech. degrees in computer science and engineering from Thapar University (2016) and Punjab Technical University (2012), respectively. He worked at Thapar University as a Lecturer from Apr`16-Oct`16. Now, he is a post-doctoral researcher in MobiSec Lab. at Department of Information Security Engineering, Soonchunhyang University, Republic of Korea. He is member of various professional bodies and past Chair for ACM Student Chapter-Patiala. He has served as a member of Program Chair Committee for the Journal of Wireless Mobile Networks, Ubiquitous Computing, and Dependable Applications and various international conferences.

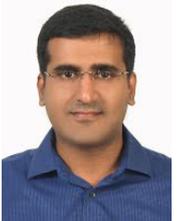

**Kathiravan Srinivasan** received his B.E., in Electronics and Communication Engineering and M.E., in Communication Systems Engineering from Anna University, Chennai, India. He also received his Ph.D., in Information and Communication Engineering from Anna University Chennai, India. He is presently working as a faculty in the department of computer science and information engineering at National Ilan University, Taiwan. He has played an active role in organizing several International Conferences, Seminars and Lectures. He has been a key note speaker in many International Conferences and IEEE events.